\documentclass[iop]{emulateapj}
\usepackage{color}
\usepackage{graphicx}
\usepackage{longtable}
\usepackage{lineno}
\usepackage{draftcopy}
\usepackage{threeparttable}
\usepackage{rotating}
\usepackage{multirow}
\slugcomment{version \today: fm}
\shorttitle{}
\shortauthors{}

\begin{document}
\title{The Extended Spatial Distribution of 
Globular Clusters in the Core of the Fornax Cluster}
\author{R.~D'Abrusco\altaffilmark{1}, M.~Cantiello\altaffilmark{2}, M.~Paolillo\altaffilmark{1}, 
V.~Pota\altaffilmark{3}, N.R.~Napolitano\altaffilmark{3}, L.~Limatola\altaffilmark{3}, 
M.~Spavone\altaffilmark{3}, A.~Grado\altaffilmark{3}, E.~Iodice\altaffilmark{3}, 
M.~Capaccioli\altaffilmark{1}, R.~Peletier\altaffilmark{4}, G.~Longo\altaffilmark{1},
M.~Hilker\altaffilmark{5}, S.~Mieske\altaffilmark{5}, E.K.~Grebel\altaffilmark{6},
T.~Lisker\altaffilmark{6}, C.~Wittmann\altaffilmark{6}, G.~van de Ven\altaffilmark{6} \& G.~Fabbiano\altaffilmark{7}}

\altaffiltext{1}{University of Naples Federico II, C.U. Monte SantAngelo, Via Cinthia, 80126, Naples, Italy}
\altaffiltext{2}{INAF-Astronomical Observatory of Teramo, Via Maggini, 64100, Teramo, Italy}
\altaffiltext{3}{INAF - Astronomical Observatory of Capodimonte, via Moiariello 16, Naples, I-80131, Italy}
\altaffiltext{4}{Kapteyn Astronomical Institute, University of Groningen, PO Box 72, 9700 AB Groningen, 
The Netherland}
\altaffiltext{5}{European Southern Observatory, Karl-Schwarzschild-Strasse 2, 85748 Garching bei 
M{\"u}nchen, Germany}
\altaffiltext{6}{Astronomisches Rechen-Institut, Zentrum f{\"u}r Astronomie der Universit{\"a}t 
Heidelberg, M{\"o}nchhofstrasse 12-14, 69120 Heidelberg, Germany}
\altaffiltext{7}{Harvard-Smithsonian Center for Astrophysics, 60 Garden Street, Cambridge, MA 02138, USA}

\begin{abstract}

We report the discovery of a complex extended density enhancement in the Globular Clusters (GCs) 
in the central $\sim\!0.5(^{\circ})^{2}$ ($\sim\!0.06$ Mpc$^{2}$) of the Fornax cluster, 
corresponding to $\sim\!50\%$ of the area within 1 core radius. This overdensity 
connects the GC system of NGC1399 to most of those of neighboring galaxies within 
$\sim\!0.6^{\circ}$ ($\sim\!210$ kpc) along the W-E direction. The asymmetric density structure suggests 
that the galaxies in the core of the Fornax cluster experienced a lively history of
interactions that have left a clear imprint on the spatial distribution of GCs. The extended 
central dominant structure is more prominent in the distribution of blue GCs, while red GCs show density 
enhancements that are more 
centrally concentrated on the host galaxies. We propose that the relatively small-scale density structures in 
the red GCs are caused by galaxy-galaxy interactions, while the extensive spatial distribution of blue GCs
is due to stripping of GCs from the halos of core massive galaxies by the Fornax gravitational potential. Our investigations is based on density maps of candidate GCs 
extracted from the multi-band VLT Survey 
Telescope (VST) survey of Fornax (FDS), identified in a three-dimensional color space and further selected based
on their $g$-band magnitude and morphology.

\end{abstract}

\keywords{galaxies: clusters: individual (Fornax) - galaxies: evolution - galaxies: individual (NGC1399)}

\section{Introduction}
\label{sec:intro}

In the hierarchical $\Lambda$CDM paradigm the most massive galaxies grow through mergers and
accretions of multiple smaller galaxies~\citep{springel2005}. These interactions leave 
their footprints in the galaxies' dynamics, kinematics, chemistry and morphology, and in particular
in galaxy halos~\citep[e.g.][]{zolotov2010} where dynamical timescales are longer than at small
galactocentric distances. In this context, historically much attention has been paid to the properties 
of the globular cluster systems (GCSs) of the galaxies~\citep{brodie2006}. In the framework
of the two-phase galaxy formation model~\citep[e.g.][]{oser2010,rodriguez2015}, GCs 
trace both the {\it in situ} star formation and the accretion phases. Recently it has been shown that 
the spatial distribution of the GC systems of massive early-type galaxies~\citep[e.g.][]{dabrusco2015,blom2014,dabrusco2014a,dabrusco2014b,dabrusco2013b} has inhomogeneities 
that may reflect the host's accretion history. This 
approach complements the study of the GCS kinematics~\citep[e.g.][]{strader2011,coccato2013,napolitano2014,pota2015a,pota2015b}.

Deep, multi-band optical imaging data covering large areas of the sky 
can be used to reconstruct the projected 2D distribution of GCs within nearby clusters
of galaxies. The spatial distribution of GCs in the Virgo cluster, for instance, has been studied using both
Sloan Digital Sky Survey~\citep{lee2010} and the Canada-France-Hawaii Telescope
data~\citep[][Next Generation Virgo Survey collaboration]{durrell2014}. In this paper we 
discuss the properties of the 2D spatial distribution of candidate GCs in the central region 
of the Fornax cluster using Fornax Deep Survey (FDS) data taken with the VST. 

\section{The Data and Catalog Extraction}
\label{sec:data}

The FDS data analyzed in this letter were acquired as Guaranteed Time Observations for the 
surveys FOrnax Cluster Ultra-deep Survey~(FOCUS; P.I. R. Peletier) and VST Early-type GAlaxy 
Survey~(VEGAS; P.I. M. Capaccioli), based on observations taken at the ESO La Silla Paranal 
Observatory. Imaging in the $u$, $g$, $r$ and $i$ filters 
was performed with the 2.6-m ESO VLT Survey Telescope (VST)
at Cerro Paranal in Chile~\citep{schipani2012}. VST is equipped with the wide field camera 
OmegaCAM~\citep{kuijken2011}, whose field of view covers $1\!\times\!1(^{\circ})^{2}$ in the 
[0.3, 1.0] $\mu$m wavelength range, with mean pixel scale of $0.21\arcsec$/pixel.

The images used in this letter cover 7 partially overlapping fields 
centered on the core of the Fornax cluster, 
for a total of $\sim\!8.4(^{\circ})^{2}$. Images were acquired using the {\it step-dither} strategy, consisting 
of a cycle of short exposures alternately centered on the field and on contiguous regions adjacent to the field. 
This approach, adopted in other photometric 
surveys~\citep[e.g.][]{ferrarese2012}, allows an accurate estimate of the sky background around bright and extended 
galaxies. For each field, we obtained 76 exposures of 150s in the $u$-band, 54s in the $g$ and $r$ 
bands and 35s in the $i$-band, for a total exposure time of 3.17, 2.25 and 1.46 hours respectively, 
reaching S/N$\sim$10 at $\!23.8,24.8,24.3,23.3$ 
magnitudes for point-like sources in the $u$, $g$, $r$ and $i$, respectively,
with no significant spatial variations. The average seeing  
and the field-to-field standard deviation are 1\farcs17$\pm$0\farcs08 in the $g$-band and 
0\farcs87$\pm$0\farcs07 in the $r$ band. Other bands display comparable variations 
over the observed fields.

The image reduction 
was performed with the VST-Tube imaging pipeline~(see~\citealt{capaccioli2015} 
and~\citealt{iodice2016} for descriptions of the pipeline and the observing strategy).
The photometric analysis of the images followed the prescriptions in~\citet{cantiello2015}.
We extracted the catalog of all sources by independently running SExtractor~\citep{bertin1996} 
in each filter. To improve the detection of sources around bright 
galaxies, we modeled and subtracted the galaxies using the IRAF \texttt{ellipse} routine and ran 
SExtractor on a $6\!\times\!6(\arcmin)^{2}$ cutouts of the frame 
centered on the extended sources. We obtained aperture magnitudes within 
a 8-pixel diameter ($\sim1\farcs68$ at OmegaCAM resolution), and applied aperture
corrections to infinite radius calculated with the growth curves of 
bright isolated point-like sources, independently for each band and 
pointing~\citep{cantiello2015}.

The catalogs in the four bands were matched with a $0\farcs5$ radius, larger than the
rms on the residuals of the differences between coordinates of overlapping detections~\citep{capaccioli2015}, 
producing a matched catalog of $\sim\!8.2\cdot 10^{4}$ sources. The techniques used to 
obtain the catalog of sources will be 
described in a forthcoming paper~\citep{cantiello2016}. Corrections for Galactic foreground extinction were 
applied following~\cite{draine2003} and the~\cite{schlafly2011} reddening map.

\section{Selection of Candidate GCs}
\label{sec:selection}

We designed a new method that employs multi-wavelength photometry and a sample of 
{\it bona fide} confirmed GCs for the selection of candidate GCs. The confirmed GCs are used
to model the region of the three-dimensional (3D) color space occupied by GCs (hereinafter, 
the {\it locus}). The main steps required to extract the candidate GCs from the general catalog of 
sources are:

\begin{itemize}
	\item Select sources located within the GC 3D color space {\it locus};
	\item Apply magnitude cuts to the color-selected sources and discard extended sources; 
\end{itemize}

GCs occupy a well-defined region in the optical color space~\citep{rhode2001,pota2013}. 
We define the {\it locus} as the smallest compact 
region in the 3D space generated by the Principal Components (PCs) associated with their
optical colors $u\!-\!g$, $g\!-\!r$ and $r\!-\!i$ (zero centered and scaled to unity variance), 
containing a fixed fraction of training set GCs. Modeling the GC {\it locus} in the PC space ensures the
simplest possible geometry by virtue of the definition of 
PCs itself~\cite[see][]{dabrusco2013b}. Only photometric sources detected in all 
bands located within the PC space {\it locus} are considered. Moreover, we select only 
sources whose ``color-error ellipsoid" (ellipsoids with semi-axes equal to the PC-transformed 
uncertainties on each color) intersects the {\it locus} for $\geq50\%$ of its volume. 

\begin{figure*}[h]
	\includegraphics[width=0.9\textwidth,angle=0]{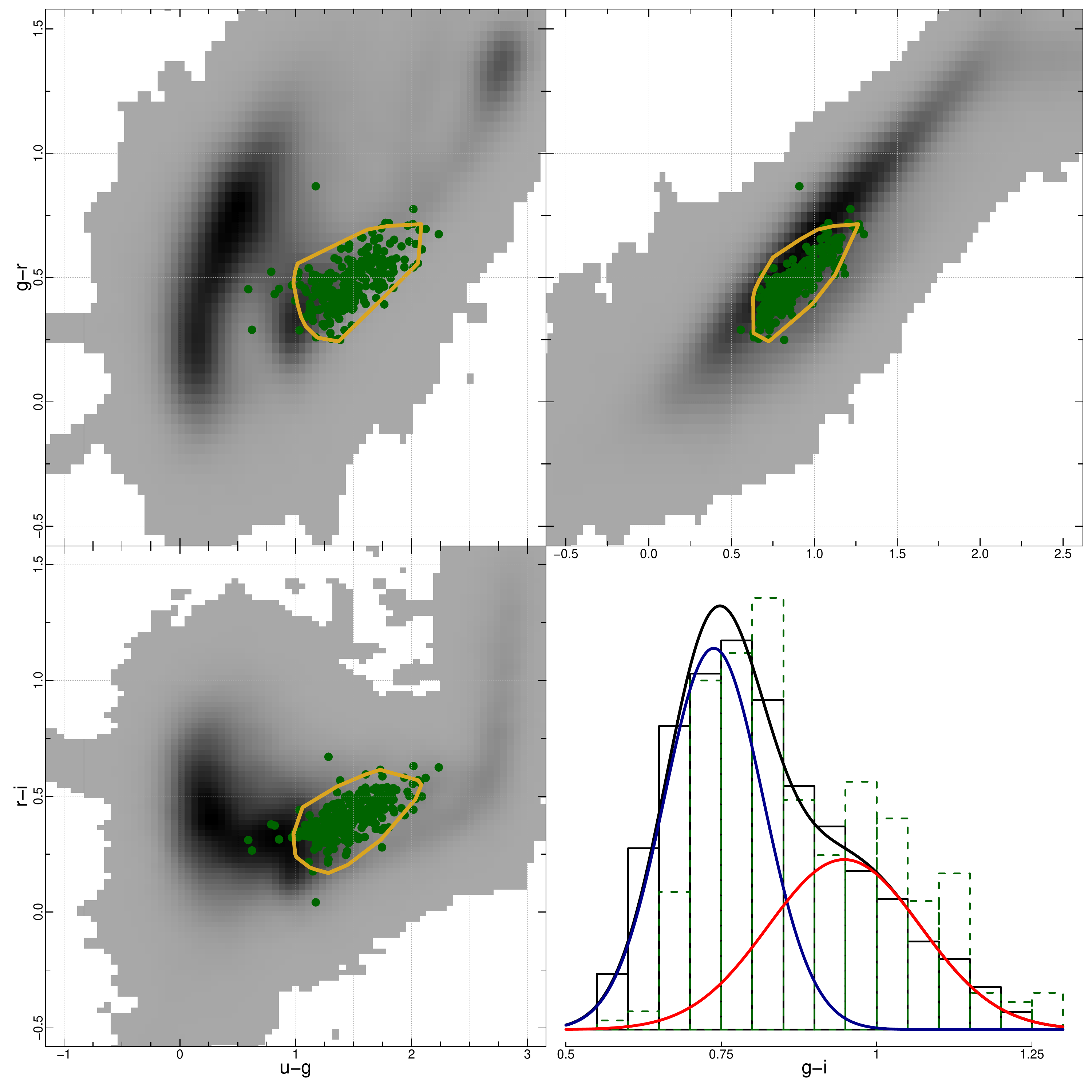}	
	\caption{Top and bottom left: projections of the PC color space {\it locus} model on three 
	color-color diagrams (yellow lines). The FDS counterparts of 
	the~\cite{schuberth2010} training set (green symbols) and the gray-scale density map
	of the general catalog of photometric sources are also shown. Bottom right: rescaled histograms of the 
	$g\!-\!i$ color of all candidate GCs (black), the training set (dashed) and the
	Gaussians fitting the red and blue subclasses.}
	\label{fig:gc_selection}
\end{figure*}

The training set used in this letter is the sample of NGC1399 
spectroscopically confirmed GCs~\citep[][A and B]{schuberth2010}, covering $\sim\!0.1(^{\circ})^{2}$ 
around the host. The GCs {\it locus} 
contains all the contiguous cells of a regular lattice with at least one training set member. 
The number of cells of the lattice, i.e. the spatial resolution of the grid along each
PC axis is set such that the average density per 
cell over the region of the PC space considered is $\sim\!1$.
Empty cells surrounded by cells containing members of the training set are included in the {\it locus} 
to obtain a simply connected geometry. Figure~\ref{fig:gc_selection} shows the projections of the 
PC space GC {\it locus} on three color-color diagrams.

We apply a brightness selection on the $g$-band magnitude and a size selection to the color-selected 
sources. We exclude sources brighter than $m_{g}\!=\!20$ to 
avoid contamination by stars and ultra-compact dwarf galaxies~\citep[cp.][]{mieske2008}. Sources fainter 
than $m_{g}\!=\!23$, the faintest $g$-band magnitude of 
the FDS counterparts of the training set GCs, are also discarded. Since GCs at the distance of the 
Fornax cluster appear mostly as unresolved sources with the VST spatial resolution, 
we employ the SExtractor $CLASS\_STAR$ parameter to select star-like sources\footnote{The 
CLASS\_STAR parameter is estimated by a neural network trained on 10 photometric parameters 
extracted from simulated images. The extension index produced by the neural network 
ranges between 0 and 1~\citep{bertin1996}.}. 
Based on the distribution of the~\cite{schuberth2010} GCs, we required 
that candidate GCs have $CLASS\_STAR\!\geq\!0.3$. We checked that our selection is consistent 
with the different technique based on the position of the sources in the $m_{g}^{(8\mathrm{pxl})}$ vs 
$\Delta m_{g}\!=\!m_{g}^{(8\mathrm{pxl})}\!-m_{g}^{(4\mathrm{pxl})}$ plane~\citep[see][]{jennings2014}, 
where the $g$-band magnitudes are measured within fixed apertures of 4 and 8 pixels. 
We found differences for less than $3\%$ of the candidate GCs selected.

\begin{table*}
	\centering	
	\caption{Left) Loadings, standard deviations and 
	variance fractions of the PCs associated with the colors of the training set; Mid) Lower and upper 
	boundaries for all parameters used for the selection; Right) Number 
	of red, blue GCs and parameters of the best-fit Gaussians.}
	\begin{tabular}{lccc|lcc|lcc}
	\tableline
		&{\small PC$_{1}$}	&{\small PC$_{2}$}& {\small PC$_{3}$}&	{\small Parameter} 	& {\small Lower} 	& {\small Upper}	&	& {\small Blue}	&	{\small Red} \\
		&		&		&		&					   					 & {\small boundary}	& {\small boundary}	&	&		&		\\
	\tableline
{\small $u\!-\!g$}	&0.579	&0.555	&0.597	&{\small PC$_{1}$}	& 	-1.56		& 1.92	& {\small N$_{GCs}$}	& $\sim\!1.8\!\cdot\!10^{3}$	& $\sim\!1.1\!\cdot\!10^{3}$	\\
{\small $g\!-\!r$}		&0.551	&-0.806	&0.215	&{\small PC$_{2}$}	& 	-0.47		& 0.41	& {\small $\mu_{(g\!-\!i)}$}	&	0.74	& 0.95\\
{\small $r\!-\!i$}		&0.597	& 0.215	&-0.773 	&{\small PC$_{3}$}	& 	-0.12		& 0.11 	&{\small $\sigma_{(g\!-\!i)}$}	& 0.08& 0.12\\
{\small $\sigma$}	&1.66	&0.49	&0.13	&{\small $m_{g}$}	& 	20		&	23	&			&		&\\
{\small $\sigma^{2}$(\%)}	&0.71	&	0.18	&	0.11	&{\scriptsize CLASS\_STAR}& 	0.3		&	-	&			&		&\\
	\tableline
	\label{tab:table}
	\end{tabular}		
\end{table*}

We required the PC {\it locus} to contain 95\% 
of the training set. The ``observed'' recovery rate is slightly smaller ($\sim\!93\%$)
because of the selections on the $g$-band magnitude and ``CLASS\_STAR" parameter.
We estimate the background of our selection as the density of all sources 
in two regions of $\sim\!0.25(^{\circ})^2$ located along the borders of the observed frame and 
devoid of bright galaxies. Since these regions are placed within 1 Fornax virial radius~\citep{drinkwater2001}, 
the background population is likely composed of both 
contaminants (stars and galaxies) and GCs. The average total density of background sources
is 0.05$\pm$0.01 objects arcmin$^{-2}$ (0.03$\pm$0.01 and 0.02$\pm$0.01 selected as blue and 
red candidate GCs respectively).
%Since we cannot exclude the presence of {\it bona fide} GCs in these regions located within 1 virial 
%radius~\citep{drinkwater2001}, though, this approach is likely to overestimate the contamination.

\section{The Catalog of Candidate GCs}
\label{sec:catalog}

The catalog of candidate GCs contains $\sim\!3000$ sources. Using the
Gaussian Mixture Modeling code~\citep[GMM,][]{muratov2010}, we determined that the candidate GCs
$g\!-\!i$ histogram is bimodal (as expected given the training set color distribution). 
The best-fit parameter of the two Gaussians 
are shown in Table~\ref{tab:table} (right). We adopted the $g\!-\!i\!=\!0.85$ threshold to separate red from 
blue candidate GCs, yielding $\sim\!1095$ red candidate GCs ($\sim\!37\%$ of the total) 
and $\sim\!1853$ blue candidate GCs ($\sim\!63\%$). A fixed color threshold is employed because 
we focus on the intracluster population, whose properties are considered to be independent of 
individual galaxies. Hence, 
the variations of the GC color distribution as function of the distance from the centers of the host 
galaxies~\citep{kim2013,cantiello2015} and the differences of the red/blue bimodality of the individual 
GCSs of~\citep[cp.][]{jordan2015} are not considered.

Using a sample of GCs extracted from Hubble data around NGC1399~\citep{puzia2014}, 
we find that the completeness magnitude of our population of 
candidate GCs is $\sim\!22.5$, 1.5 magnitudes brighter than the average $g$-band turnover 
of the GCLF $m^{(g)}_{TO}\!=\!24$~\citep{villegas2010,jordan2007}. While 
the $g\!-\!i$ distribution (bottom-right panel in Figure~\ref{fig:gc_selection}) is consistent with those 
in~\cite{bassino2006} and~\cite{kim2013}, our catalog
is dominated by the blue component of the GCs population because the FDS 
data, unlike the~\cite{schuberth2010} training set, 
cover a wide area of the cluster at large distances from the hosts, 
where GCs are overwhelmingly blue~\citep[cp.][]{durrell2014}.
The crossmatch of our catalog with the deeper ($m_{V}^{\mathrm{lim}}\!=\!23.7$)~\cite{kim2013} sample 
of $UBVI$-photometry selected candidate GCs in a $36\arcmin\!\times\!36\arcmin$ region around NGC1399 
returns $\sim70\%$ of common GCs. We also recover $\sim\!85\%$ of the kinematically 
confirmed GCs from~\cite{bergond2007}.

\section{Maps of the Spatial Distribution of GCs}
\label{sec:results}

We determined the density maps of the spatial distribution of candidate GCs applying the K-Nearest 
Neighbor (KNN) method~\citep[see][]{dressler1980,dabrusco2013a} on a regular grid covering the 
observed region. KNN density, defined as $d_{K}\!=\!K/(\pi\cdot r^{2}_{K})$, 
is directly proportional to the neighbor index $K$ and inversely proportional to the 
area of the circle defined by the distance of the $K$-th closest source to the center of the grid 
cells, where the density is calculated. The parameter $K$ was varied over 
the interval $[3, 9]$~\citep[see][for a discussion on the choice of $K$]{dabrusco2015}. 
This strategy provides a non-zero density in each cell of the grid, including those not 
containing candidate GCs. $K$ is a measure of the richness and size expressed in number 
of members, of the structures to which the density map is sensitive to. Maps with small $K$ are 
dominated by compact density structures, while large $K$'s highlight 
more extended structures. In the following, we will discuss $K\!=\!9$ maps to 
focus on the features of the distribution of candidate GCs on large spatial scales.
Since background contribution to the catalog of candidate GCs is only known statistically, 
the shapes and areas of the density structures discussed henceforth are to be considered 
approximations.

The significance of the density structures is calculated as the complement to the probability of 
the number of candidate GCs observed within the structure of being caused by fluctuations of the 
background population, which we assume to follow the Poisson statistics. 
For density structures located within other overdensities, the significance has been determined 
relative to poissonian fluctuations of a population with density equal to the  
density of all candidate GCs in the underlying density structure. By assuming that 
background sources are independent, our estimates are lower limits to the real 
significances of the density structures.
In the case of enhancements including compact GCSs, the contribution of the hosts' GC population is 
accounted for by removing all the sources within a circle containing 95\% of the GCS population centered 
on each galaxy. We assume a power-law radial density profile with slope from the literature when available or
fixed to $\alpha\!=\!-2$ otherwise.

\begin{figure*}[h]
	\includegraphics[width=0.9\textwidth,angle=0]{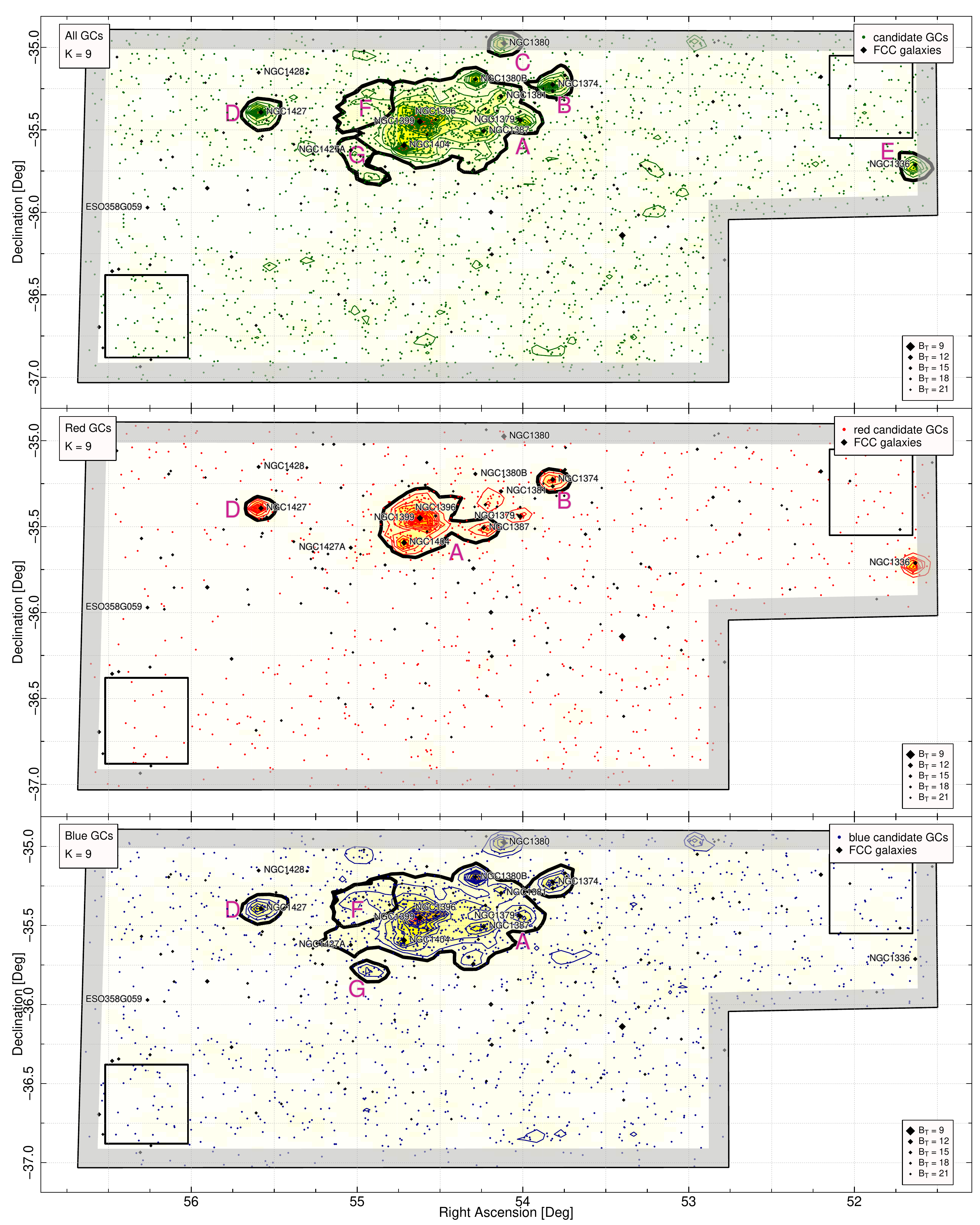}
	\caption{Top to bottom: $K\!=\!9$ density maps of the spatial distribution of all, 
	red and blue candidate GCs. All structures discussed in the text are highlighted and  
	labeled. All maps display contours 
	at ten log-spaced densities levels (normalized to peak) starting at 
	0.05, FCC galaxies (diamonds) size-coded according to their $B_{T}$ magnitude 
	and the background regions. Density in the shaded region may be underestimated because of
	border effects.}
	\label{fig:densitymaps}
\end{figure*}

Figure~\ref{fig:densitymaps} shows the density maps for (top to bottom) all, red and blue
candidate GCs. The distribution of all candidate GCs is dominated by a large, irregularly shaped 
overdensity roughly centered on NGC1399 and extended along the W-E axis, within a region of 
$\sim\!0.7(^{\circ})^{2}$ (A). The significance of this structure is $>\!99.9\%$. This structure stretches out
to connect NGC1399 with most of the nearby galaxies (NGC1404, NGC1387, 
NGC1396, NGC1380B, NGC1381), and contains $\sim\!30\%$ of all candidate GCs, 
($35\%/65\%$ red/blue), corresponding to $\sim\!19\%$ of total number of 
blue candidate GCs and $\sim\!9\%$ of total number of red candidate GCs in our catalog.

Other galaxies associated with noticeable overdensities are NGC1374 (B), 
NGC1380 (C), NGC1427 (D) and NGC1336 (E)
in agreement with earlier works on their GCSs~\citep{kisslerpatig1997a,kisslerpatig1997b}, while ESO358G059 
and NGC1428 lack visible density enhancements. Two spatial features not associated
to bright galaxies are clearly visible in the density map of all candidate GCs: a large enhancement of circular 
shape (F, $>\!84.2\%$) $\sim\!0.5^{\circ}$ NE
of NGC1399; and a hook-shaped structure (G, $>\!91.6\%$) at $\sim\!0.25^{\circ}$ SE of the 
NGC1404 overdensity and seemingly 
connected to the structure A through the NGC1427A density enhancement.
Outside the core region, excluding the structures of the 
NGC1427, NGC1336 and NGC1380 GCSs, the distribution of candidate GCs is featureless and 
homogeneous on spatial scales $\geq\!0.25^{\circ}$, except for a 
$\sim\!0.35\!\times\!0.35(^{\circ})^{2}$ region centered on R.A.$\approx\!55.6^{\circ}$ and 
Dec$\approx\!-36^{\circ}$ almost devoid of candidate GCs ($>\!90.5\%$). We checked the photometric quality 
in this area, but did not notice differences with the other regions of the observed frame.

The central overdensity associated with NGC1399 is 
significantly more compact for red than for blue GCs. Although the accuracy of the 
density maps deteriorates with smaller sample size, this difference is still evident for brighter 
magnitude-selected subsets of red and blue candidate GCs. The structures F and G are not visible in the density 
map of red candidate GCs but are still detected in the blue GC density map (with $>\!99.9\%$ and $>\!98.5\%$
significance, respectively), hinting to a possible accretion-related origin~\citep[e.g.][]{cote1998}.
The notable differences of the NGC1380 and NGC1336 density structures for red and blue candidate 
GCs are caused by the combined effects of the fixed $g\!-\!i$ color 
threshold, the low number of 
candidate GCs due to high galaxy background levels, and the relative shallowness of the $u$-band images 
that favors the selection of blue candidate GCs. Moreover, near the edges
of the surveyed region the KNN with $K\!=\!9$ may underestimate the density of compact, numerically small 
or extended, only partially covered GC subpopulations.

\begin{figure*}[h]
	\includegraphics[width=0.8\textwidth,angle=0]{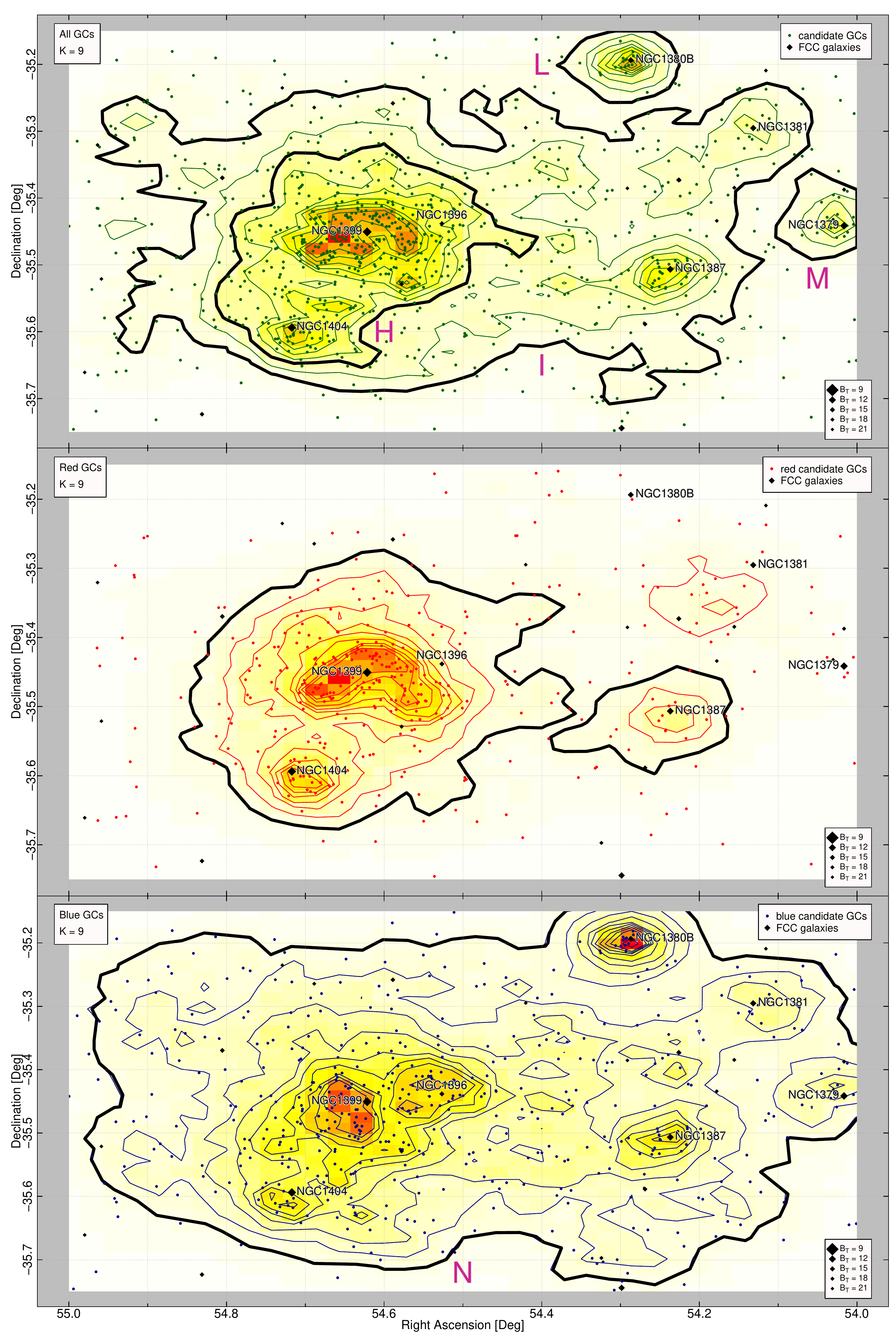}
	\caption{As in Figure~\ref{fig:densitymaps}, zoomed in on the core of the Fornax cluster.}
	\label{fig:densitymaps_zoom}
\end{figure*}

We produced additional density maps for the cluster core, using a finer grid. 
Figure~\ref{fig:densitymaps_zoom} shows the maps for all, red and blue candidate GCs in a 
$\sim\!1^{\circ}\mathrm{(R.A.)}\times\!\sim\!0.5^{\circ}\mathrm{(Dec.)}$
box roughly centered on NGC1399. The strong overdensity connecting NGC1399
and NGC1404 along the SE-NW direction (H, $>\!92.1\%$), has peaks associated 
with the galaxies centers. The investigation of the morphology 
of the NGC1399-NGC1404 complex, interpreted by~\cite{bekki2003} with NGC1399 stripping GCs from NGC1404, 
requires the modeling of the radial profiles of these GCSs and is 
not discussed here. While~\cite{kim2013} reported a $0\farcm5$ displacement between the 
optical center of NGC1399 and the center of its GC distribution, attributed to 
recent interactions, our analysis of the spatial distribution of candidate GCs in the central region of NGC1399
is constrained by the completeness of our catalog of candidate GCs. Using samples of {\it bona fide} HST GCs 
in the central region of NGC1399~\citep{puzia2014,jordan2015}, we find that the completeness of 
our catalog of candidate GCs increases from a minimum of $\sim\!10\%$ within 0\farcm5 from the center of the 
galaxy, corresponding to a $g$-band surface brightness 
$\sim\!17\ \mathrm{mag}/(\arcsec)^{2}$, to $\sim\!55\%$ at 
1\farcm5 ($\sim\!23\ \mathrm{mag}/(\arcsec)^{2}$), and remains constant between $\sim50\%$ and $\sim55\%$ 
with no evident trend up to $\sim6\farcm5$ ($\sim\!23\ \mathrm{mag}/(\arcsec)^{2}$). For this reason, we 
cannot determine the nature of the deficit 
observed in Figure~\ref{fig:densitymaps_zoom}. Comparable completeness is measured near the centers 
of NGC1380, NGC1336 and NGC1387.

On large spatial scales, the main NGC1399-NGC1404 overdensity extends to the 
W joining with the density enhancements associated with NGC1387 and NGC1381 (I, $>\!99.9\%$). 
The bridge connecting the NGC1399 overdensity to the NGC1387 GCS has been 
reported by~\cite{bassino2006} in the spatial distribution of blue candidate GCs within $\sim\!1(^{\circ})^{2}$  
around NGC1399 and, 
independently, by~\cite{kim2013}. The NGC1380B (L) and 
NGC1379 (M) density enhancements are isolated. 
The density map of red candidate GCs (Figure~\ref{fig:densitymaps_zoom}, mid) still shows a large 
overdensity in the core of the Fornax cluster composed by two sub-concentrations 
centered on NGC1399 and NGC1404, connected by a lower density valley. The other bright galaxies in the 
field are not associated with clear density enhancements. 
In the density maps of blue candidate GCs, the central structure (N, $>\!99.9\%$) is similar to
structure I, but also includes the NGC1380B and NGC1379 enhancements.

\section{Discussion}
\label{sec:conclusions}

We report a density structure in the spatial distribution of candidate GCs in a region $\sim\!0.5(^{\circ})^{2}$ 
within the core of the Fornax cluster of galaxies, connecting NGC1399 to the surrounding galaxies 
NGC1404, NGC1387, NGC1381 and NGC1380B. Our findings expand on 
the discovery by~\cite{bassino2006} of a GC bridge between NGC1399 and NGC1404. 
We used $\sim\!3000$ candidate GCs extracted from VST $ugri$ images covering the central 
$\sim\!8.4(^{\circ})^{2}$ of the cluster. The method employed for the selection of candidate 
GCs relies on the modeling of the {\it locus} occupied by confirmed GCs in the 3D color space 
generated by $ugri$-bands photometry, and the application of cuts on $g$-band magnitude 
and morphology.

\begin{figure*}[h]
	\includegraphics[width=\textwidth]{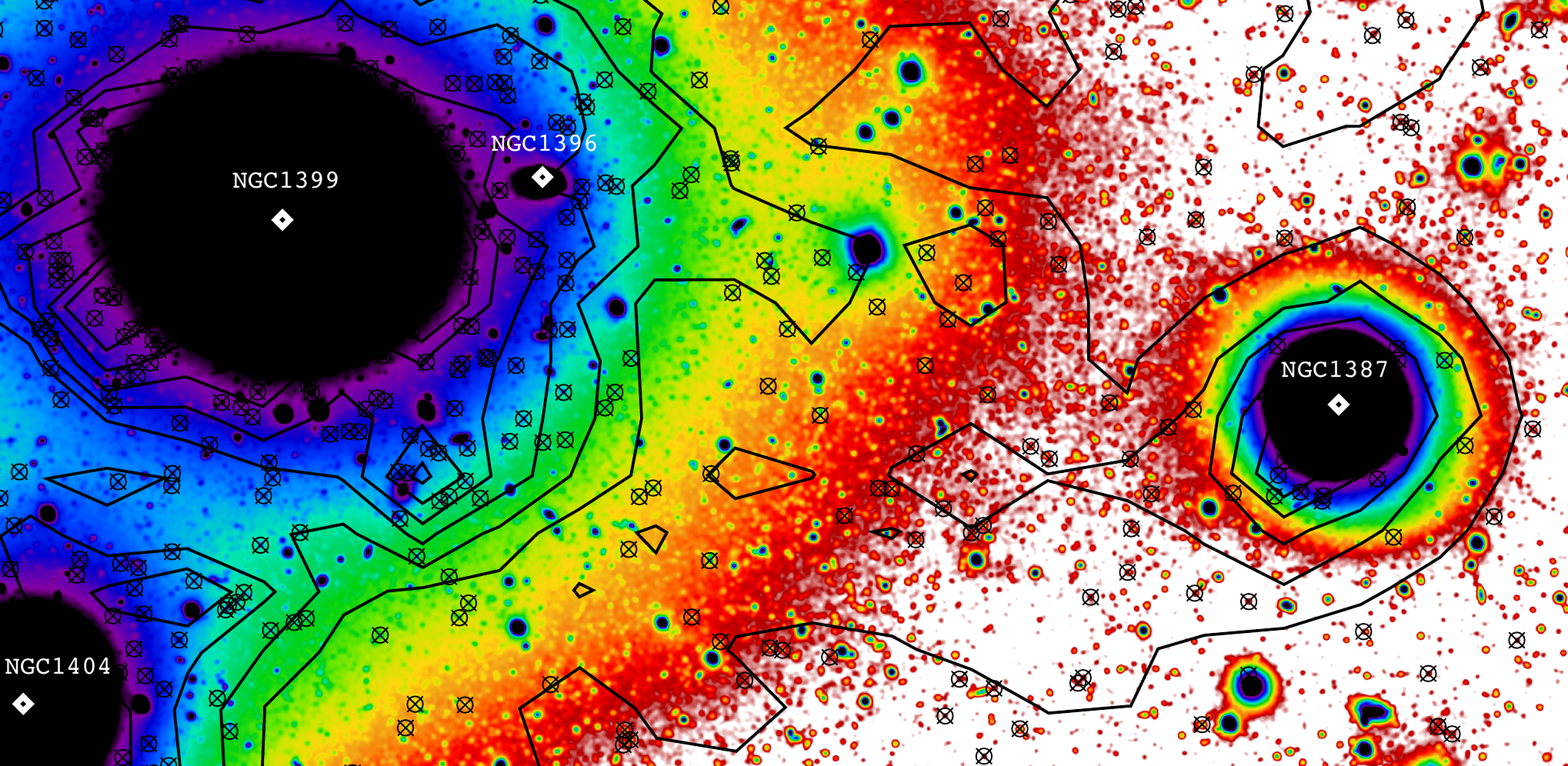}
	\caption{$K\!=\!9$ isodensity contours and all candidate GCs overplotted to the FDS $g$-band 
	image of the NGC1399-NGC1387 region~\citep{iodice2016}.}
	\label{fig:streamer}
\end{figure*}

The dominant GC overdensity displays an intricate 
morphology and is elongated on the WE direction. A similar asymmetry was observed, on scales of 
$\sim\!100$ kpc, in the NGC1399 X-ray halo emission~\citep{paolillo2002}, due to the presence of three
distinct components with different centers and sizes. The diffuse spatial structures not associated with 
galaxies suggest a long and active history of interactions which have shaped the spatial distribution of 
GCs. We re-observe the GC bridge connecting the NGC1399-NGC1404 complex to NGC1387~\citep{bassino2006,kim2013},
that seems to support the~\cite{iodice2016} claim of a low surface brightness stream in the 
FDS $g$-band light distribution (Figure~\ref{fig:streamer}). This evidence
suggests that the more massive NGC1399 may have stripped GCs from NGC1387. 
We also highlighted two isolated density structures (F and G) with complex shapes mostly 
formed by blue GCs, whose origins could be related to accretion or stripping events. 
Other isolated galaxies (NGC1380, NGC1427, 
NGC1379 and NGC1336) are associated with distinct, compact enhancements. On larger 
spatial scales we observe a featureless distribution of candidate GCs, barring few compact GCSs and 
one underdensity.

The zoomed in density map of blue GCs (Figure~\ref{fig:densitymaps_zoom}, bottom) shows that NGC1381, 
NGC1379 and NGC1380B are also connected to the main overdensity centered on NGC1399. This result
indicates that GC stripping may not have been limited to NGC1387, as proposed by~\cite{bassino2006}. 
Since the shapes of the density enhancements in the map of red GCs, barring the evident central overdensities
associated to the GCSs of the single galaxies, are reminiscent of the structure 
linking the NGC1399-NGC1404 region to NGC1387, one may speculate 
that these galaxies have experienced strong interactions which may have disrupted the GCSs of nearby satellites 
and/or gravitationally trapped relatively red intra-cluster GCs. The large-scale morphology of
blue GCs observed in both the general and the zoomed density maps may be explained with a different
mechanism, namely the trapping in the potential well in the Fornax core of mostly blue 
GCs stripped by the tidal field of the cluster from hosts at small cluster-centric 
distances, as predicted by simulations~\citep{ramos2015}.

Our conclusions are constrained by the limits of the training set and the depth of the photometric data.
As the survey nears completion, the new data acquired will be used to improve our analysis and 
provide a detailed assessment of the performance of our selection 
using a larger, more 
homogenous sample of spectroscopically selected~(P.I. Capaccioli, ID 094.B-0687;~\citealt{pota2016}) 
GCs covering $1(^{\circ})^{2}$ around NGC1399, observed  
in the context of the FDS project. Additional insights into the properties of the spatial distribution of 
GCs in Fornax will be obtained by comparing the features of the GC structures with the
phase-space properties of the~\cite{pota2016} sample, and by modeling the distinct 
spatial components of the candidate GCs population~\citep{dabrusco2015,dabrusco2014a,dabrusco2014b,dabrusco2013a,dabrusco2013b}.

{}
\end{document}